\pdfoutput=0
\documentclass[11pt]{article}
\usepackage{axodraw}
\usepackage{epsfig}
\usepackage{amsfonts}
\usepackage{amsmath}
\usepackage{bm,bbm}
\usepackage{cite}
\allowdisplaybreaks[1]

\input pix.sty

\newcommand{\la}[1]{\label{#1}}
\newcommand{\be}{\begin{equation}}
\newcommand{\ee}{\end{equation}}
\newcommand{\ba}{\begin{eqnarray}}
\newcommand{\ea}{\end{eqnarray}}
\newcommand{\rmi}[1]{{\mbox{\scriptsize #1}}}
\newcommand{\fig}{Fig.~}

\newcommand{\eq}{Eq.~}
\newcommand{\eqs}{Eqs.~}
\newcommand{\se}{Sec.~}

\newcommand{\nr}[1]{(\ref{#1})}

\newcommand{\nn}{\nonumber \\}

\renewcommand{\vec}[1]{{\bf #1}}

\renewcommand{\eq}{eq.~}
\renewcommand{\eqs}{eqs.~}
\renewcommand{\se}{sec.~}

\renewcommand{\fig}{fig.~}

\newcommand{\alphas}{\alpha_{\rm s}}
\renewcommand{\B}{\rmii{$B$}}

\newcommand{\F}{\rmii{$F$}}

\newcommand{\mpl}{m_\rmi{pl}}

\newcommand{\Nc}{N_{\rm c}}

\newcommand{\rmO}{{\mathcal{O}}}

\def\lsi{\raise0.3ex\hbox{$<$\kern-0.75em\raise-1.1ex\hbox{$\sim$}}}
\def\gsi{\raise0.3ex\hbox{$>$\kern-0.75em\raise-1.1ex\hbox{$\sim$}}}
\newcommand{\lsim}{\mathop{\lsi}}

\newcommand{\nF}{n_\rmii{F}} 
\newcommand{\nB}{n_\rmii{B}} 
\newcommand{\rmii}[1]{{\mbox{\tiny\rm{#1}}}}

\newcommand{\im}{\mathop{\mbox{Im}}}

\newcommand{\Tint}[1]{{\hbox{$\sum$}\!\!\!\!\!\!\!\int\,}_{\!\!\!\!\raise-0.9ex\hbox{$\scriptstyle{#1}$}}}
\newcommand{\Tinti}[1]{{{\Sigma}\!\!\!\!\raise0.3ex\hbox{$\int$}_\rmii{${#1}$}}}

\newcommand{\bi}{\begin{itemize}}
\newcommand{\ei}{\end{itemize}}
\newcommand{\hide}[1]{ }
\newcommand{\blind}[1]{\fbox{$ ? $}} 

\newcommand{\deltabar}{\raise-0.02em\hbox{$\bar{}$}\hspace*{-0.8mm}{\delta}}
\newcommand{\momdif}{\zeta}

\renewcommand{\P}{\mathcal{P}}

\newcommand{\X}{\mathcal{X}}
\newcommand{\aS}{\varphi} 

\newcommand{\mS}{m_\aS}

\makeatletter \@addtoreset{equation}{section} \makeatother
\renewcommand{\theequation}{\arabic{section}.\arabic{equation}}
\makeatletter
\renewcommand\section{\@startsection {section}{1}{\z@}%
                                   {-5.5ex \@plus -1ex \@minus -.2ex}
                                   {2.3ex \@plus.2ex}%
                                   {\normalfont\large\bfseries}}
\renewcommand\subsection{\@startsection{subsection}{2}{\z@}%
                                     {-3.25ex\@plus -1ex \@minus -.2ex}%
                                     {1.5ex \@plus .2ex}%
                                     {\normalfont\normalsize\bfseries}}
\renewcommand\thesection {\@arabic\c@section}
\renewcommand\thesubsection   {\thesection.\@arabic\c@subsection}
\renewcommand{\@seccntformat}[1]{%
\csname the#1\endcsname.\hspace{1.0em}}
\makeatother

\begin{document}

\flushbottom

\begin{titlepage}

\begin{flushright}
April 2023
\end{flushright}
\begin{centering}
\vfill

{\Large{\bf
  Langevin simulation of dark matter kinetic equilibration
}} 

\vspace{0.8cm}

Seyong Kim$^\rmi{a,b}_{ }$
and
M.~Laine$^\rmi{b}_{ }$

\vspace{0.8cm}

$^\rmi{a}$%
{\em 
Department of Physics, 
Sejong University, 
Seoul 143-747, Korea
\\}

\vspace*{0.3cm}

$^\rmi{b}$%
{\em
AEC, 
Institute for Theoretical Physics, 
University of Bern, \\ 
Sidlerstrasse 5, CH-3012 Bern, Switzerland \\}

\vspace*{0.8cm}

\mbox{\bf Abstract}
 
\end{centering}

\vspace*{0.3cm}
 
\noindent
Recently it has been questioned, notably in the context of the scalar
singlet dark matter model with $\mS^{ }\simeq 60$~GeV, how efficiently
kinetic equilibrium is maintained if freeze-out dynamics is pushed
down to low temperatures by
resonant effects. We outline how Langevin simulations can be employed
for addressing the non-equilibrium momentum distribution of
non-relativistic particles in a cosmological background. 
For a scalar singlet mass $\mS^{ }\simeq 60$~GeV, 
these simulations suggest that kinetic equilibrium is
a good approximation down to $T \sim 1$~GeV, with the deviation first
manifesting itself as a red-tilted spectrum. This reduces the
annihilation cross section, confirming findings from other methods
that a somewhat larger ($ < 20$\%) coupling than in equilibrium is
needed for obtaining the correct abundance.

\vfill


\end{titlepage}

\tableofcontents

%
\section{Introduction}
\la{se:intro}

Determining the abundance of dark matter in a given model requires
ingredients from general relativity, quantum field theory, and 
non-equilibrium statistical physics. While the first two are 
well-established frameworks, non-equilibrium statistical physics
is an open construction, with even the specification of the 
state hard to achieve in full generality. Therefore
it is not rare that cosmological computations  
call for physical intuition. 

Often, a fruitful approach is to estimate the rates at which various
processes take place, and then divide the variables 
into two classes, the fast and slow ones. A tractable problem is found
if the fast variables undergo many interactions within the
observation period; they can then be assumed to thermalize, 
constituting a heat bath. The slow variables remain out of 
equilibrium, but since there are fewer of them, the problem
is easier to handle.  

In dark matter computations, we normally assume that all Standard
Model particles are fast variables. For dark matter particles, the 
fastest processes are soft elastic scatterings, which may be
assumed to decohere the system. Subsequently we can focus on classical
notions, like the momentum distribution of the 
dark matter particles (``kinetic non-equilibrium''), 
and their overall number density (``chemical non-equilibrium''). 

Adjusting the momentum distribution towards the 
thermal (Bose-Einstein or Fermi-Dirac) form only requires elastic scatterings. 
In the non-relativistic regime, 
these are much faster than inelastic ones. 
Therefore, it is often a good assumption 
to impose kinetic equilibrium
from the outset, and only focus on deviations from chemical 
equilibrium.

Recently, however, the validity of this picture
has been questioned. In particular, 
the example of the scalar singlet dark matter model
(cf.,\ e.g.,\ refs.~\cite{s1,s2,s3,s4} and references therein)
has been intensively discussed in the 
non-relativistic freeze-out regime~\cite{tb,kk,drake,abe,kk2}
(cf.,\ e.g.,\ refs.~\cite{other0,other1,kin,other2,other3} for similar
effects in other models).
To be clear, let us remark that 
in the so-called \mbox{freeze-in} scenario, dynamics takes place
in the relativistic regime, and then there is 
in general no hierarchy between the kinetic and chemical 
equilibration rates, so that kinetic non-equilibrium is 
certainly present.

Following their use in the context 
of heavy ion collision experiments~\cite{mt}, 
we propose here to employ Langevin simulations 
for studying the efficiency of kinetic equilibration
in the non-relativistic regime. 
The effect of the fast variables is encoded in the 
values of two matching coefficients, which can be defined
and computed at the NLO~\cite{sch} or even 
at the non-perturbative level~\cite{eucl}.
Therefore Langevin simulations offer for 
a systematically improvable framework for 
studying strongly coupled systems, 
notably dark matter scattering off a 
Standard Model plasma at temperatures of a few~GeV.

This paper is organized as follows. 
We start by reviewing how the Langevin equation can be 
set up in an expanding background, 
in \se\ref{se:langevin}.
This is followed by a description of 
an algorithm for its
numerical solution,
and a summary of the corresponding 
simulation results, in \se\ref{se:simu}. 
In \se\ref{se:chem} we show how the non-equilibrium 
momentum distribution can be implemented in a freeze-out computation. 
Conclusions are collected 
in \se\ref{se:concl}, relegating the computation
of the matching coefficients for the scalar singlet model 
to appendix~A.

\section{Langevin equation in an expanding background}
\la{se:langevin}

We assume the universe to be described by 
a homogeneous, isotropic and spatially flat  
Friedmann-Lema\^itre-Robertson-Walker
background, with the metric
\be
 {\rm d}s^2_{ } = {\rm d}t^2 - a^2_{ }(t)\, {\rm d}\vec{x}^2 
 \;. \la{metric}
\ee
The physical 4-momentum of an on-shell particle is 
denoted by $p^{\mu}_{ }$.
The covariant derivative of its spatial components reads 
\be
 {p^{i}_{ }}^{ }_{;t}
 = 
 \dot{p}^i_{ } + H \, p^{i}_{ }
 \;, 
\ee
where $H \equiv \dot{a}/a$ is the Hubble rate and 
$
 \dot{p} \; \equiv \; 
 {\rm d}p / {\rm d}t
$.
Viewing $p^i_{ }$ as a slow variable, 
the Langevin equation takes the form 
\be
 \dot{p}^i_{ }\; = \; - (\eta + H) p^i_{ } + f^i_{ }
 \;, \la{langevin}
\ee
where $\eta$ is a friction (or ``drag'') coefficient and 
$f_{ }^i$ is a random force, taking care 
of detailed balance
(i.e.\ returning thermal energy to the heavy particle, 
in exchange for that lost through friction). 
The force obeys the autocorrelator
\be
 \bigl\langle\, 
  f^i_{ }(t^{ }_1) \, f^j_{ }(t^{ }_2) 
 \,\bigr\rangle 
 \; = \;
 \momdif \, \delta^{ij}_{ }\, \delta(t^{ }_1-t^{ }_2) 
 \;, \la{zeta_def}
\ee
where $\momdif$ is called the momentum diffusion coefficient. 
The constraint that the system should thermalize to 
a temperature $T$ imposes 
the fluctuation-dissipation relation 
\be
 \eta = \frac{\momdif \bigl\langle \vec{v}^2_{ } \bigr\rangle}{6 T^2_{ }}
 \;. \la{relation}
\ee
The average velocity can in turn be expressed as 
$ \bigl\langle \vec{v}^2_{ } \bigr\rangle \approx 3 T/ m_{\aS}^{ }$,  
where we have introduced the notation $m_{\aS}^{ }$ for the mass of 
a generic non-relativistic dark matter particle. 

When we implement the Langevin equation in a cosmological context, 
time and temperature are not independent variables.
If the system does not undergo phase transitions, so 
that the temperature evolves smoothly, we may take
\be
 x \; \equiv \; \ln\biggl( \frac{T^{ }_\rmi{max}}{T} \biggr)
 \;, \quad
 (...)' \; \equiv \; \frac{{\rm d}(...)}{{\rm d}x}
 \;, \la{x_def}
\ee
as a time-like variable
(we choose $T^{ }_\rmi{max} \equiv$ 5~GeV). 
The Jacobian to physical time is 
\be
 \frac{{\rm d} x}{{\rm d}t} = 3 c_s^2 H
 \;, 
\ee
where $c_s^2 = \partial p / \partial e$ is the speed of sound squared. 
Furthermore the entropy density, $s$, satisfies 
$
 \dot{s} + 3 H s = 0
$, 
and consequently 
$s a^3 = $ const.
If we now define dimensionless momenta as 
\be
 \hat{p}^i_{ }\; \equiv \; \frac{p^i_{ }}{s^{1/3}_{ }}
 \;, 
\ee
and denote 
\be
 \hat{\eta} \; \equiv \; \frac{\eta}{3c_s^2 H}
 \;, \quad
 \hat{\momdif} \; \equiv \; \frac{\momdif}{3c_s^2 H s^{2/3}_{ }} 
 \;, \la{hat_gamma}
\ee
then Langevin dynamics can be expressed as  
\be
 ({\hat{p}_{ }^{i}})' \; = \; 
 -\hat{\eta}\, \hat{p}^i_{ }+ \hat{f}^i_{ }
 \;, \quad
 \bigl\langle\, 
   \hat{f}^i_{ }(x^{ }_1) \, \hat{f}^j_{ }(x^{ }_2) 
 \,\bigr\rangle
  \; = \; 
  \hat{\momdif} \, \delta^{ij}_{ }\, \delta(x^{ }_1-x^{ }_2)
 \;. \la{langevin_2} 
\ee
Given the constancy of $s a^3$, 
we note that $\hat{p}^i_{ } \propto a p_{ }^i \equiv k^i_{ }$, 
known as a comoving momentum.

\begin{figure}[t]

\hspace*{-0.1cm}
\centerline{%
 \epsfysize=5.0cm\epsfbox{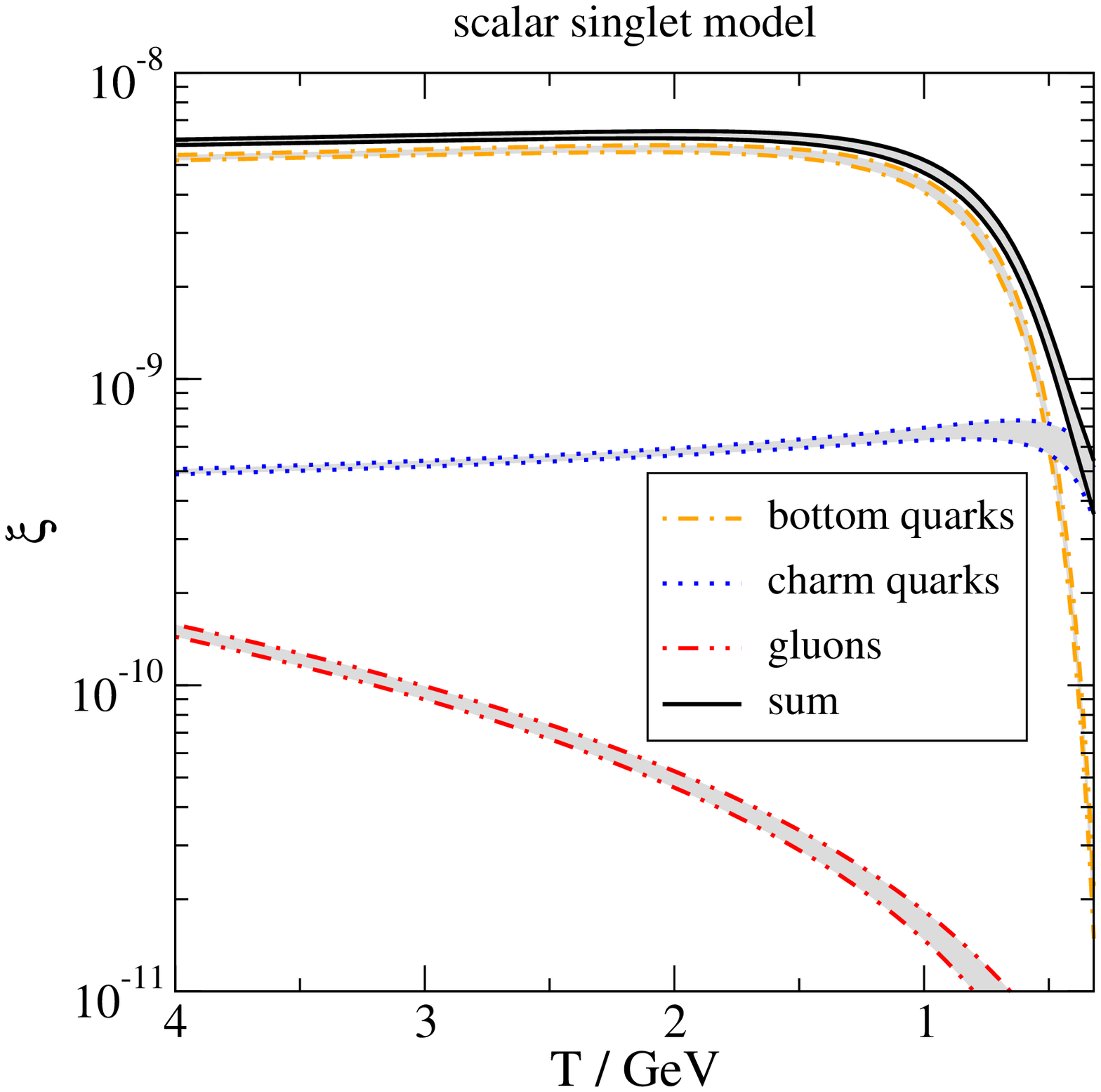}%
 \hspace{0.4cm}%
 \epsfysize=4.8cm\epsfbox{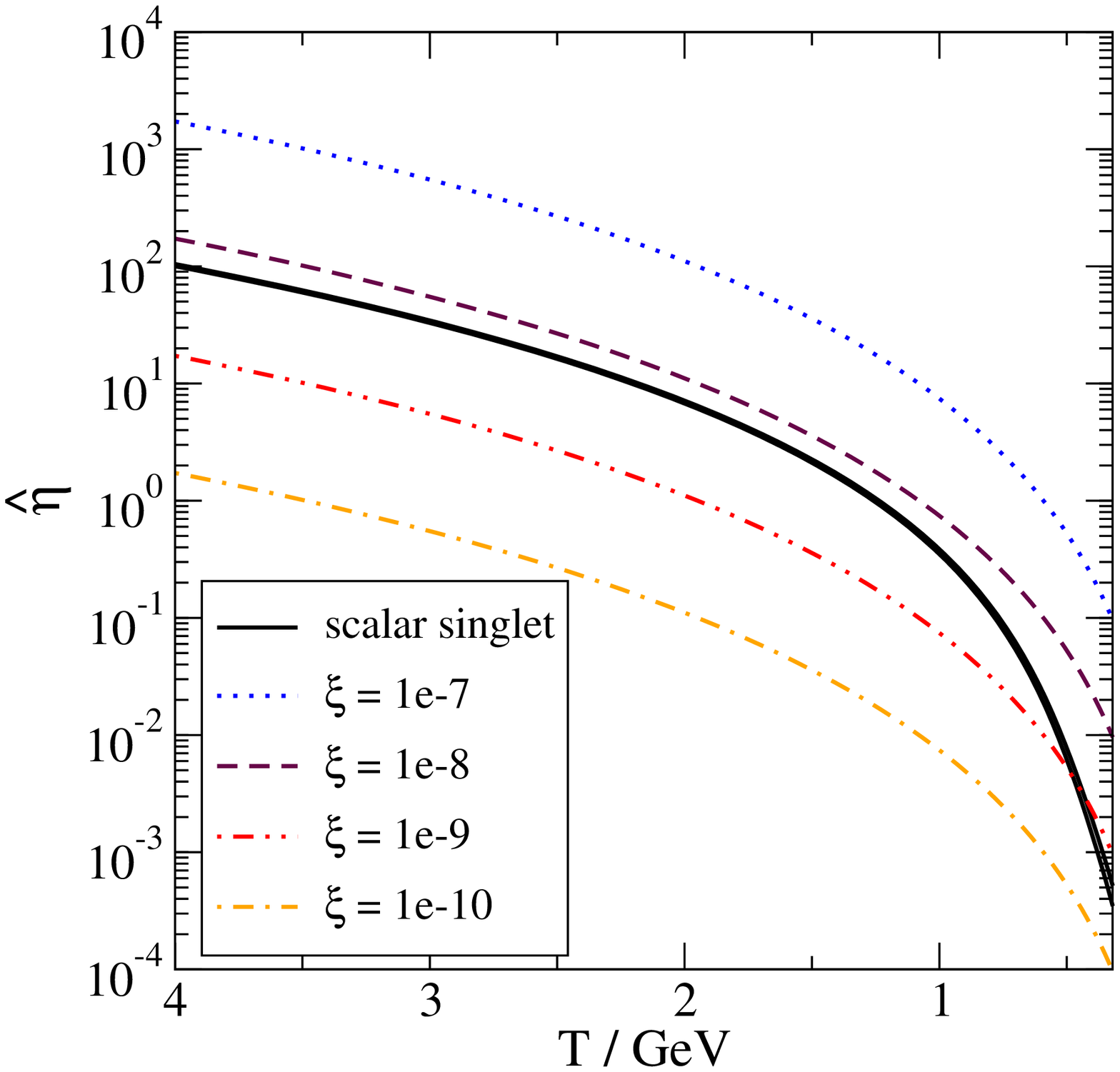}%
 \hspace{0.4cm}%
 \epsfysize=4.8cm\epsfbox{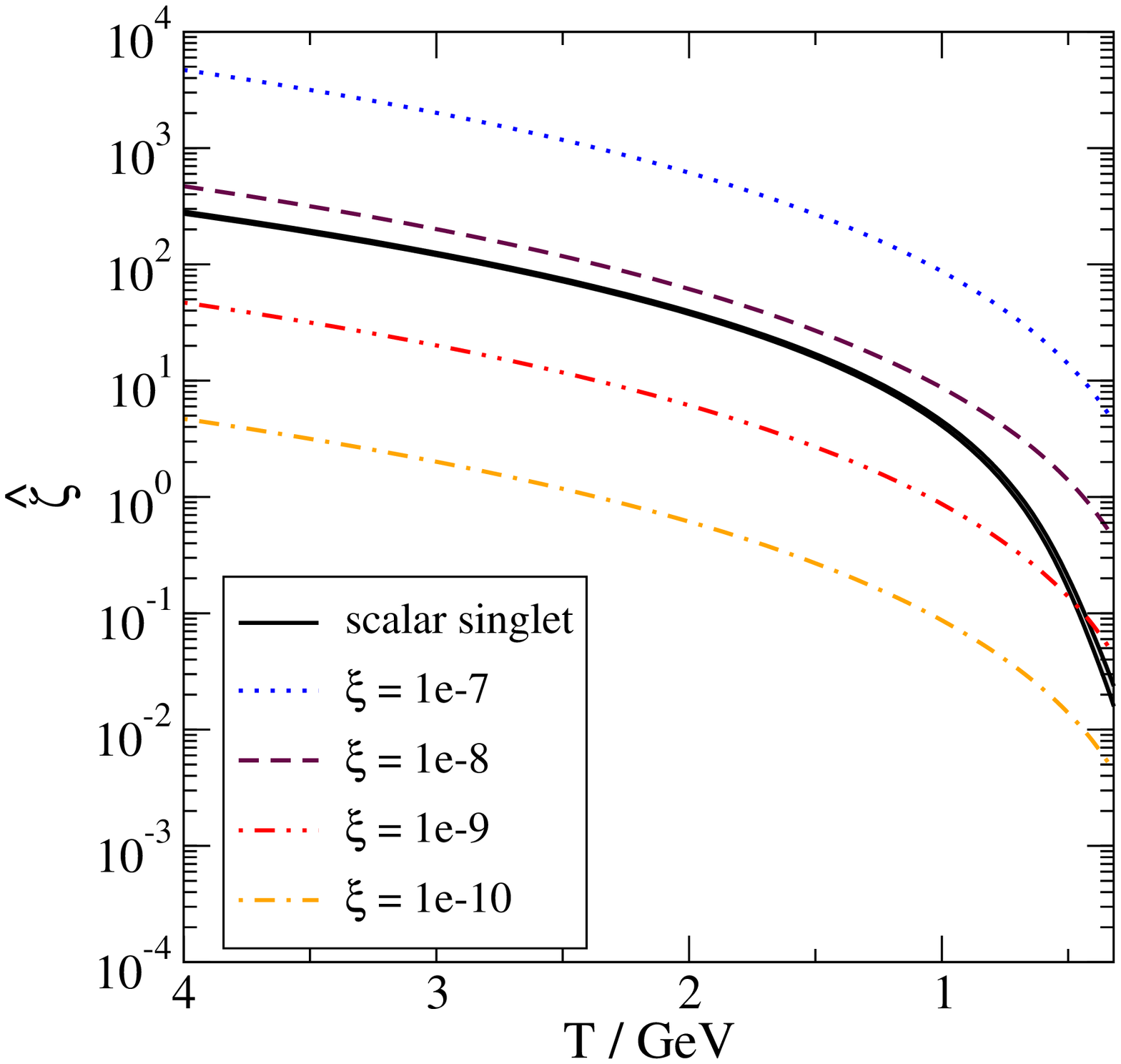}%
}

\caption[a]{\small
  Left: 
  the momentum diffusion coefficient
  obtained from \eqs\nr{gammaF_est} and \nr{gammaB_est}, 
  originating from scatterings off quarks or gluons, 
  converted into $\xi$ as defined by \eq\nr{xi}, 
  with $m^{ }_{\aS} \simeq 60$~GeV and 
  $\kappa \simeq 0.00064$~\cite{singlet_chemical}.
  Confining effects in the vicinity of the QCD crossover
  have been modelled via the substitution 
  $\Nc^{ }\to N^{ }_{\rm c,eff} < 3$~\cite{nuMSM}, 
  and 
  the difference to the tree-level value $\Nc^{ } = 3$
  has been displayed
  as a grey band. 
  Middle and right: 
  the corresponding~$\hat{\eta}$
  and~$\hat{\momdif}$
  from \eq\nr{hat_gamma}, either with $\xi$
  as displayed in the left panel (solid line), 
  or for various fixed values of $\xi$. 
  For thermodynamic
  potentials we have inserted 
  estimates from ref.~\cite{ls}, 
  tabulated at 
   {\tt http://www.laine.itp.unibe.ch/eos15/}.
  }

\la{fig:xi}
\end{figure}

A key element of the dynamics is 
that the coefficients $\hat{\eta}$ and $\hat{\momdif}$ are not
constant but evolve rapidly with $x$. The Hubble rate reads
\be
 H = \sqrt{\frac{8\pi e}{3 \mpl^{2}}}
 \;, 
\ee
where 
$
 e
$
is the energy density 
and 
$
 \mpl^{ }\approx 1.22091 \times 10^{19}_{ }
$~GeV
is the Planck mass. 
Since $e\sim T^4$ in the Standard Model plasma, $H$ scales as $\sim T^2$.
The entropy density scales as $s\sim T^3$.
The coefficient $\momdif$ is suppressed by the mass of the 
dark matter particle and that of the mediator between the visible 
and dark sectors. For dimensional reasons, we may write it as 
\be
 \momdif \; \equiv \; \frac{\xi\, T^7_{ }}{ (100~\mbox{GeV})^4_{ } }
 \;, \la{xi}
\ee
with $\xi$ displaying modest temperature dependence. 
Different contributions to $\xi$ in the scalar singlet model, 
derived in appendix~A,\footnote{%
 The computation of $\zeta$ in appendix~A amounts to the
 quantum-field theoretic evaluation of the 2-point real-time 
 correlation function of the force that changes momenta. 
 The force in turn is identified as the time derivative of
 the spatial components of the particle number current.
 Subsequently, model-dependent but weakly coupled fields
 (dark matter, mediator) can be handled perturbatively, 
 leaving over a correlation function of strongly coupled
 objects (QCD currents composed of quarks and gluons), 
 which could in principle be evaluated non-perturbatively.  
 } 
are shown 
in \fig\ref{fig:xi}(left). The speed of sound squared can often
be approximated as $3 c_s^2 \simeq 1$, though it experiences 
corrections when mass thresholds are crossed. 
Putting all of these scalings together, we expect 
$
 \hat{\eta}
 \propto
 (T / \mbox{GeV})^4_{ }
$
and 
$
 \hat{\momdif}
 \propto
 (T / \mbox{GeV})^3_{ }
$
(cf.\ \fig\ref{fig:xi}(middle, right)).
This implies that kinetic equilibrium
is likely to be lost at low~$T$. 

Equation~\nr{langevin_2} is a linear inhomogeneous 
first order differential equation, and as such it can be 
given an explicit formal solution, 
\be
 \hat{p}^i_{ }(x^{ }_2) 
 \; = \; 
 \hat{p}^i_{ }(x^{ }_1) 
 \, \exp \biggl[\, 
  - \int_{x^{ }_1}^{x^{ }_2}
    \! {\rm d}y\, \hat{\eta}(y)
 \,\biggr]
 + 
 \int_{x^{ }_1}^{x^{ }_2}
 \! {\rm d}z \, 
 \hat{f}^i_{ }(z) \, 
 \exp\biggl[\, 
    \int_{x^{ }_2}^z \! {\rm d}y \, \hat{\eta}(y)
 \,\biggr]
 \;. \la{soln_1} 
\ee
Taking an average over the noise, moments can be obtained, 
for instance
\be
 \bigl\langle \, \hat{\vec{p}}^2_{ }(x^{ }_2) \, \bigr\rangle
 \; = \; 
 \bigl\langle \, \hat{\vec{p}}^2_{ }(x^{ }_1) \, \bigr\rangle
 \, \exp \biggl[\, 
  - 2 \int_{x^{ }_1}^{x^{ }_2}
      \! {\rm d}y\, \hat{\eta}(y)
 \,\biggr]
 \; + \; 
 3 \int_{x^{ }_1}^{x^{ }_2}
 \! {\rm d}z \, 
 \hat{\momdif}(z) \, 
 \exp\biggl[\, 
    2 \int_{x^{ }_2}^z \! {\rm d}y \, \hat{\eta}(y)
 \,\biggr]
 \;. \la{soln_2} 
\ee
A numerical illustration is shown in \fig\ref{fig:pp}.
Compared with such moments, 
the advantage of a direct numerical solution of 
\eq\nr{langevin_2} is that all moments can be 
obtained at once, from the momentum distribution.

\begin{figure}[t]

\hspace*{-0.1cm}
\centerline{%
 \hspace{0.4cm}%
 \epsfysize=7.3cm\epsfbox{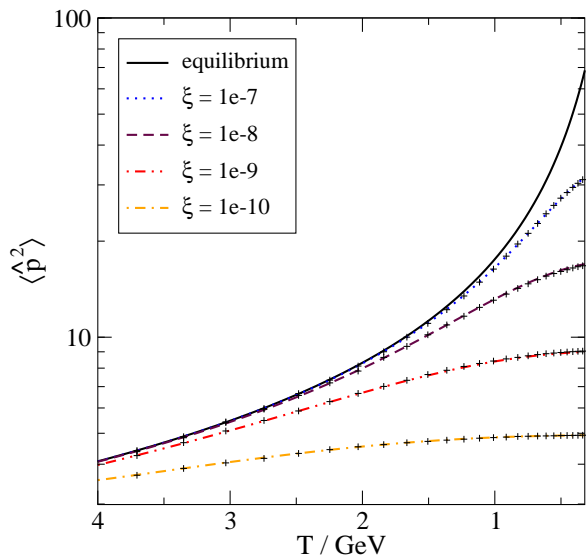}%
}

\caption[a]{\small
  The average rescaled momentum squared, from \eq\nr{soln_2}
  (dashed and dotted lines), 
  compared with the equilibrium value, 
  from \eq\nr{soln_3} (solid black line), 
  and from fits based on \eq\nr{P_noneq} to Langevin 
  simulation data (crosses). 
  The system started
  from an equilibrium configuration at $T=5.0$~GeV.
  The larger $\xi$ (cf.\ \eq\nr{xi}), the longer
  the system stays close to equilibrium. 
  }

\la{fig:pp}
\end{figure}

Finally we note that the would-be equilibrium distribution, 
at any given temperature, is obtained by {\em assuming} the 
coefficients temperature-independent, so that there is a lot of 
time for the system to adjust to the given situation. 
If we normalize the momentum 
distribution in analogy with cosmological power spectra, so that 
\be
 \int \! {\rm d}\ln(\hat{p}) \, \P^{ }(\hat{p}) = 1
 \;, \quad
 \hat{p} \; \equiv \; |\hat{\vec{p}}|
 \;,  
\ee
then the equilibrium form reads
\be
 \P^{ }_\rmi{eq} \; = \; 
 \frac{ \hat{p}^3 }{2\pi^2}
 \biggl( \frac{4\pi\hat{\eta}}{\hat{\momdif}} \biggr)^{3/2}_{ }
 \exp\biggl(  - \frac{ \hat{p}^2 \hat{\eta} }{ \hat{\momdif} } \biggr) 
 \;. \la{P_eq}
\ee
To see how fast the system approaches this limit, 
we can make the approximation of
temperature-independent coefficients in \eq\nr{soln_2}, 
obtaining
\be
 \bigl\langle \, \hat{\vec{p}}^2_{ }(x^{ }_2) \, \bigr\rangle^{ }
 \; \stackrel{\rmi{eq}}{\approx} \; 
 \biggl[\, 
 \bigl\langle \, \hat{\vec{p}}^2_{ }(x^{ }_1) \, \bigr\rangle
   - 
  \frac{3\hat{\momdif}}{2\hat{\eta}}
 \,\biggr]
 \, e_{ }^{\, 
        -2 \hat{\eta} (x^{ }_2 - x^{ }_1)
           \,}
 \; + \; 
  \frac{3\hat{\momdif}}{2\hat{\eta}}
 \;. \la{soln_3}
\ee
For $\hat\eta (x^{ }_2 - x^{ }_1) \ll 1$
the terms $ 3\hat\momdif / (2\hat\eta) $  cancel, 
so that  
non-equilibrium manifests itself
by the system staying
close to the old value.
For $\hat\eta (x^{ }_2 - x^{ }_1) \gg 1$, 
the system loses memory of initial conditions
and moves towards 
the equilibrium value 
$\langle \hat{\vec{p}}^{2}_{ }\rangle^{ }_\rmi{eq}
 = 3\hat\momdif / (2\hat\eta) \sim m^{ }_\aS / T $. 
As the variable $x$ is of $\rmO(1)$, 
we can say that kinetic decoupling
starts when $\hat{\eta} < 1$.

%
\section{Time discretization and numerical simulations}
\la{se:simu} 

We now move on to a numerical integration of \eq\nr{langevin_2}.
The problem is technically non-challenging 
(unless one is interested in the distribution of momenta
in the far UV tail), 
and we employ a simple-minded approach. 
The time-like variable $x$ is discretized, and we denote by 
$\hat{p}^i_n$, $\hat{\eta}^{ }_n$ and $\hat{\momdif}^{ }_n$ the 
values of the momenta and coefficients on the corresponding grid. 
For \eq\nr{langevin_2} we use the Ito discretization with Gaussian noise,
\begin{equation}
  \hat{p}^i_{n+1} = \hat{p}^i_n - \hat{\eta}{ }_n \, \hat{p}^i_n {\rm d}x
  + \hat{f}^i_n \sqrt{{\rm d}x}
  \;, \quad 
  \langle\, \hat{f}^i_n \hat{f}^j_m \,\rangle = 
  \hat{\momdif}^{ }_n \, \delta^{ij}_{ } \, \delta^{ }_{m n}
  \;. \la{langevin_discrete}
\end{equation}
Here, the thermodynamic functions appearing
in $\hat{\eta}^{ }_n$ and $\hat{\momdif}^{ }_n$ are
interpolated from the tabulated values given in \cite{ls}
by the cubic-spline method \cite{recipes}. 

As the coefficients in \eq\nr{langevin_discrete}
change by 4 orders of magnitude
in the temperature range studied
(cf.\ \fig\ref{fig:xi}), it is important to have
a small enough time step.
We have found that this
requirement can be sufficiently satisfied with ${\rm d}x = 10^{-6}_{ }$.
The initial $\hat{p}^i_{ }$'s
are drawn from the equilibrium distribution at $T = 5$ GeV. The
momentum distribution at each $x$ is obtained by histograms 
produced from $N = 10^{5}_{ }$ independent runs. 
The error in each histogram bin is calculated from a jackknife
analysis, with a block size of $10^3_{ }$.

If the momentum distribution obtained from the simulation
is denoted by $\P$, then a useful observable is its ratio
to the would-be equilibrium value from \eq\nr{P_eq}, 
\be
 r \; \equiv \; \frac{\P}{\P^{ }_\rmii{eq}}
 \;. \la{def_r}
\ee
Snapshots of $r$ as obtained from the simulations
are illustrated in \fig\ref{fig:tabulate}.
For $\xi$, we consider a number of fixed values  
(cf.\ \fig\ref{fig:xi}), spanning the range that is realistic
for the model considered in appendix~A.
It is clear from \fig\ref{fig:tabulate} that the
system rapidly departs from equilibrium if $\xi$ is small, 
and that it does so by retaining power at small momenta, 
as a remnant from an earlier time 
(cf.\ the discussion around \eq\nr{soln_3}).

\begin{figure}[t]

\hspace*{-0.1cm}
\centerline{%
 \hspace{-0.4cm}%
 \epsfysize=5.3cm\epsfbox{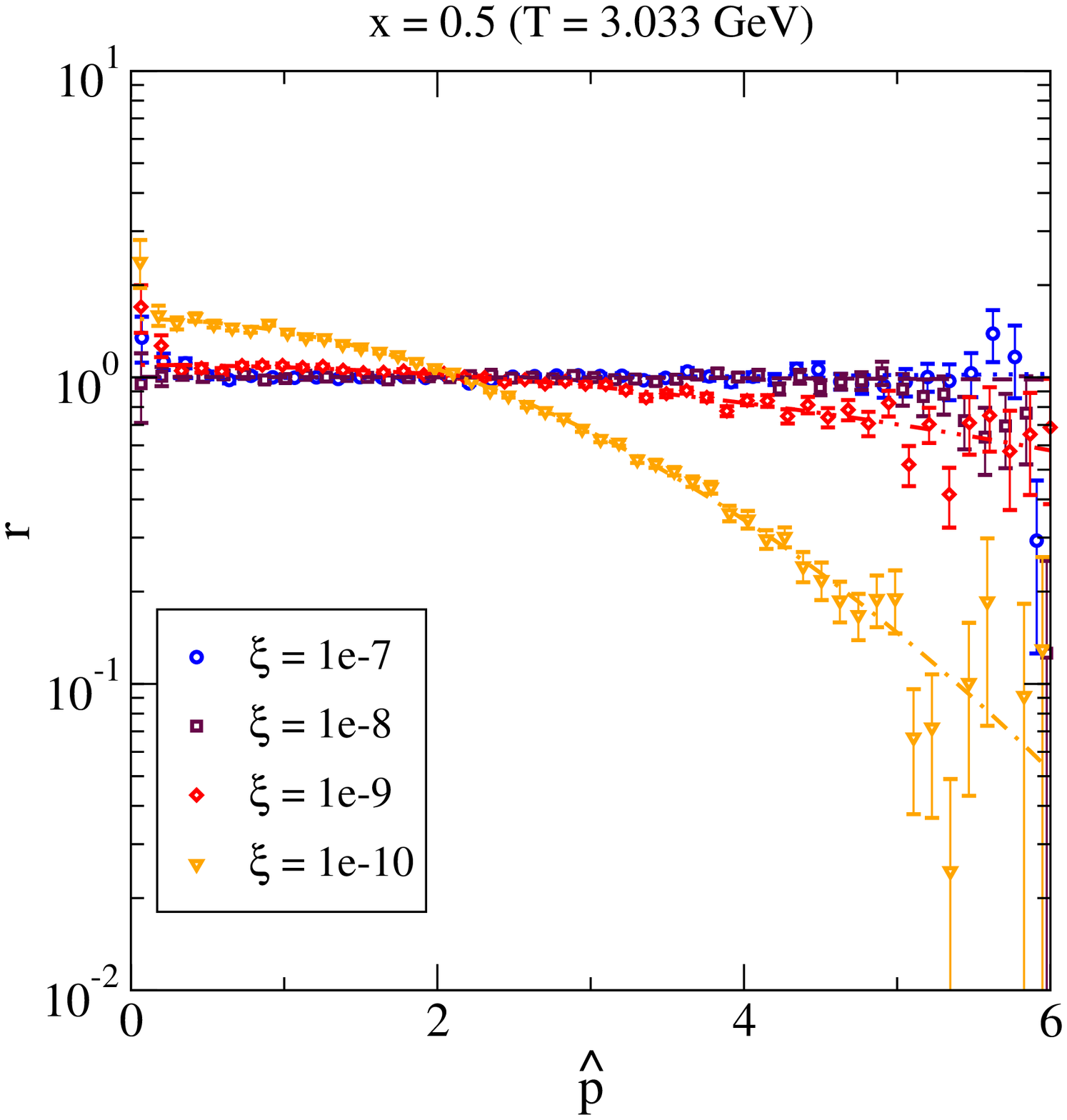}%
 \hspace{0.4cm}%
 \epsfysize=5.3cm\epsfbox{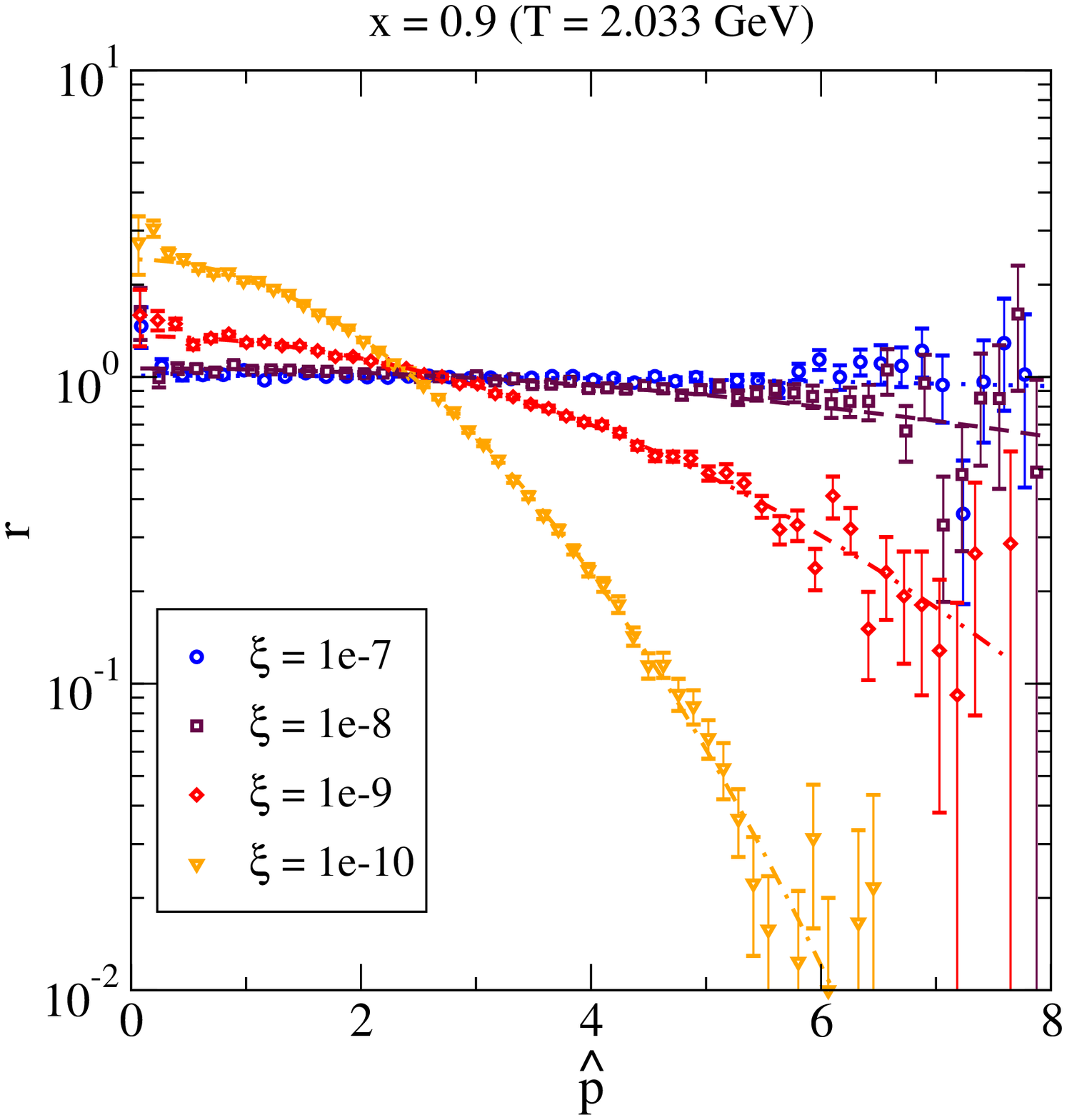}%
 \hspace{0.4cm}%
 \epsfysize=5.3cm\epsfbox{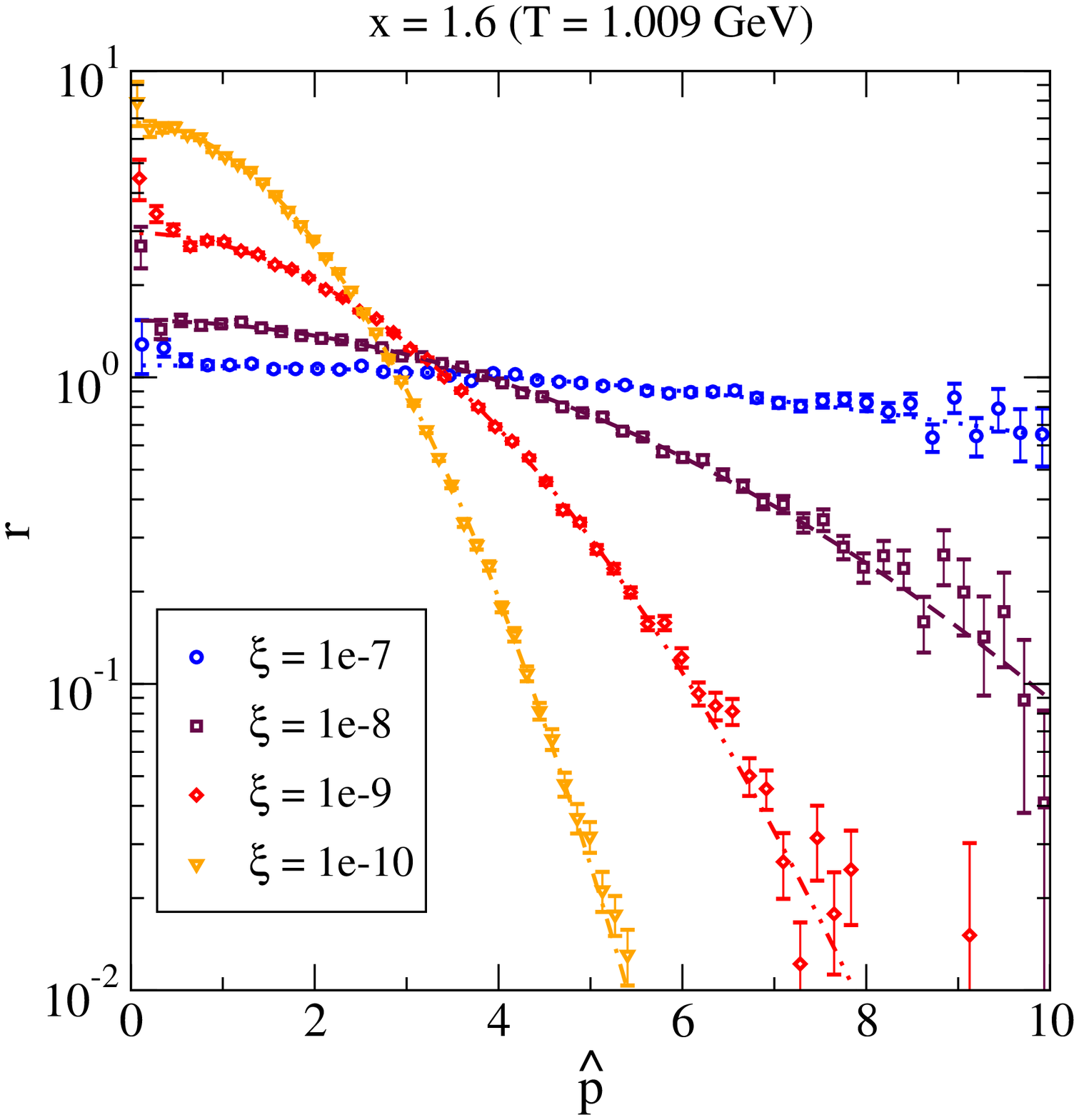}%
}

\caption[a]{\small
  Examples of the non-equilibrium modification of the momentum 
  distribution, denoted by~$r$ (cf.\ \eq\nr{def_r}), 
  for various temperatures and values of $\xi$ (data points), 
  compared with fits to \eq\nr{P_noneq} (lines). The good 
  performance of the fits confirms that the momentum distribution
  maintains a Gaussian form 
  even after the system falls out of equilibrium. 
  }

\la{fig:tabulate}
\end{figure}

We find that the simulation results are well represented by the 
functional form of \eq\nr{P_eq}, parametrized however by 
a different coefficient which we denote by $\alpha$, 
\be
 \P \; \simeq \;
   \frac{4\hat{p}^3
   \alpha^{3/2}_{ } \exp\bigl( -\alpha \hat{p}^2 \bigr)
   }{\sqrt{\pi}}
 \;, \quad
 \bigl\langle \hat{\vec{p}}^2_{ } \bigr\rangle
  \; \simeq \; \frac{3}{2\alpha}
 \;, \quad
   \alpha \; \neq \; \frac{\hat{\eta}}{\hat{\momdif}}
 \;. \la{P_noneq}
\ee
The corresponding fits are illustrated in \fig\ref{fig:tabulate}.
These results lead to an easy crosscheck of the accuracy of 
the solution, as illustrated by the crosses in \fig\ref{fig:pp}.
This implies that the information in \fig\ref{fig:pp}, 
originating from \eq\nr{soln_2}, 
is sufficient for determining the full momentum distribution.
 The conclusion that kinetic non-equilibrium is captured by the  
 value $\langle \hat{\mathbf{p}}^2 \rangle$ is not new
 but appears frequently in the literature, however we have arrived
 at it as a result of a systematic computation, 
 rather than adopted it as a starting point.

%
\section{Chemical equilibration with non-equilibrium momentum distribution}
\la{se:chem}

Having determined the non-equilibrium momentum distribution
in \se\ref{se:simu}, the next step is to implement it in 
the equation governing the dark matter number density. 
For the scalar singlet model in the resonant regime, 
it was demonstrated in ref.~\cite{singlet_chemical} 
that inelastic processes are to a very good 
approximation described by the leading-order 
$
 \varphi\varphi \leftrightarrow h
$
reaction, where~$h$ stands for the Standard Model Higgs boson, 
set on-shell. 
Noting furthermore that freeze-out physics 
takes place deep in the 
non-relativistic regime, where $\pi T \ll m^{ }_\aS$, 
we may employ the Boltzmann form for the equilibrium
distribution function. We denote the non-equilibrium
phase space distribution by $f^{ }_\aS$ and 
the equilibrium one by 
$
 \bar{f}^{ }_\aS \equiv \exp({-\epsilon^{ }_\aS/ T}_{ })
$.
The Boltzmann equation for $f^{ }_{\aS}$ then takes the form
\be
 \bigl( \partial^{ }_t - H p^{ }_1 \partial^{ }_{p^{ }_1} \bigr)
 f^{ }_{\aS^{ }_1}
 \;\approx\;
 -
  \int_{\vec{p}^{ }_2,\vec{p}^{ }_h}
  \hspace*{-4mm}
  \frac{\kappa^2 v^2 (2\pi)^4_{ }
        \delta(\epsilon^{ }_{\aS_1} + \epsilon^{ }_{\aS_2} - \epsilon^{ }_h)
        \delta^{(3)}_{ }(\vec{p}^{ }_1 + \vec{p}^{ }_2 - \vec{p}^{ }_h)
  }{
   8 \epsilon^{ }_{\aS_1} \epsilon^{ }_{\aS_2} \epsilon^{ }_{h}
  }
  \, 
  \bigl(
    f^{ }_{\aS_1^{ }} 
    f^{ }_{\aS_2^{ }} 
   - 
    \bar{f}^{ }_{\aS_1^{ }} 
    \bar{f}^{ }_{\aS_2^{ }} 
  \bigr)
 \;, \la{boltzmann}
\ee
where 
$
 \epsilon^{ }_{\aS^{ }_i} 
 \equiv \sqrt{p_i^2 + m_{\aS}^2}
$, 
$ 
 p^{ }_i \equiv |\vec{p}^{ }_i|
$, 
$
 \int_{\vec{p}^{ }_i} 
 \equiv \int \! \frac{{\rm d}^3\vec{p}^{ }_i}{(2\pi)^3_{ }}
$, 
and unspecified notation 
is explained in the context of \eq\nr{L}.

To proceed, we integrate \eq\nr{boltzmann} over $\vec{p}^{ }_1$. 
The number density is denoted by 
$
 n^{ }_{\aS} \equiv \int^{ }_{\vec{p}^{ }_i} f^{ }_{\aS^{ }_i}
$, 
and the left-hand side becomes 
$
 (\partial^{ }_t + 3 H ) n^{ }_\aS
$
after partial integration. 
The system is closed by noting that on the right-hand side, 
we may parametrize the non-equilibrium distribution function
with the information obtained in \se\ref{se:simu}, as
\be
 f^{ }_{\aS^{ }_i} \; \equiv \; 
 \frac{n^{ }_\aS}{\bar{n}^{ }_\aS} 
 \, r(\hat{p}^{ }_i) 
 \, \bar{f}^{ }_{\aS^{ }_i}
 \;, \quad 
 i = 1,2
 \;, 
\ee
with the constraints that 
$
 \int_{\vec{p}^{ }_i} r(\hat{p}^{ }_i) \bar{f}^{ }_{\aS^{ }_i} 
 = 
 \int_{\vec{p}^{ }_i} \bar{f}^{ }_{\aS^{ }_i} 
 = 
 \bar{n}^{ }_{\aS}
 = \frac{ m_\aS^2 T }{2\pi^2}
 K^{ }_2\bigl( \frac{m^{ }_\aS}{T} \bigr)
$, 
where $K^{ }_2$ denotes a modified Bessel function.
The constraint on $r$ follows from the definition in \eq\nr{def_r}. 
Going over to the variables $Y^{ }_\aS \equiv n^{ }_\aS / s$, 
$\bar{Y}^{ }_\aS \equiv \bar{n}^{ }_\aS / s$,
and again replacing time through $x$ from \eq\nr{x_def}, 
we can rewrite the evolution equation as 
\be
 \partial^{ }_x Y^{ }_{\aS} 
 \; \approx \; 
 - \frac{s}{3 c_s^2 H}
   \,
   \bigl[\, 
     \langle \sigma v^{ }_\rmi{rel} \rangle 
     \,  Y_\aS^2 
     - 
     \langle \overline{ \sigma v }^{ }_\rmi{rel}  \rangle
     \,  \bar{Y}_\aS^2  
   \,\bigr]
 \;. \la{lw}
\ee
Here the first annihilation cross section is averaged over the 
non-equilibrium momentum distribution, 
\ba
 \langle \sigma v^{ }_\rmi{rel} \rangle 
 & \equiv & 
 \int_{ \vec{p}^{ }_1,\vec{p}^{ }_2 }
 \hspace*{-3mm}
 \frac{\pi\kappa^2 v^2
 \delta\bigl(\, 
         \epsilon^{ }_{\aS^{ }_1} + \epsilon^{ }_{\aS^{ }_2} 
         - \sqrt{
              (\vec{p}^{ }_1 + \vec{p}^{ }_2)^2_{ } + m_h^2 
                }
       \,\bigr)
  }{
    4 \epsilon^{ }_{\aS^{ }_1} \epsilon^{ }_{\aS^{ }_2}
      (\epsilon^{ }_{\aS^{ }_1} + \epsilon^{ }_{\aS^{ }_2})
 }
      \, 
   \frac{ 
       \bar{f}^{ }_{\aS^{ }_1}
       \bar{f}^{ }_{\aS^{ }_2}
       r(\hat{p}^{ }_1)
       \,r(\hat{p}^{ }_2)
        }{ \bar{n}^2_\aS }
 \nn[2mm] 
 & = & 
 \frac{\kappa^2 v^2}{32\pi^3 \bar{n}^2_\aS }
 \int_{m^{ }_\aS}^{\infty} 
 \! {\rm d}\epsilon^{ }_{\aS^{ }_1} \,
 \int_{
  \epsilon^{-}_{\aS^{ }_2}
      } 
     ^{ 
  \epsilon^{+}_{\aS^{ }_2}
      }
   {\rm d}\epsilon^{ }_{\aS^{ }_2}
      \, 
       \bar{f}^{ }_{\aS^{ }_1}
       \bar{f}^{ }_{\aS^{ }_2}
       \,r\Bigl( 
           {\textstyle
              \frac{\sqrt{\epsilon^2_{\aS_1} - m_\aS^2 }}{s^{1/3}}
           }
          \Bigr)
       \,r\Bigr(
           {\textstyle
              \frac{\sqrt{\epsilon^2_{\aS_2} - m_\aS^2 }}{s^{1/3}}
           }
          \Bigr)
 \;, \la{sigmav}
\ea
where the integration bounds can be established as 
\be
 \epsilon^{\pm}_{\aS^{ }_2}
 \; \equiv \; 
    \biggl( \frac{m_h^2}{2 m_\aS^2} - 1\biggr) \epsilon^{ }_{\aS^{ }_1}  
    \pm
    \frac{m_h }{m^{ }_\aS}
    \sqrt{\Bigl( \frac{m_h^2}{4 m_\aS^2} - 1 \Bigr) 
          \bigl( \epsilon_{\aS^{ }_1}^2 - m_\aS^2 \bigr) }
 \;. 
\ee
The second term in \eq\nr{lw} contains an average
with respect to equilibrium distributions, 
\be
 \langle \overline{ \sigma v }^{ }_\rmi{rel}  \rangle
  \; \equiv \;    
 \langle \sigma v^{ }_\rmi{rel} \rangle 
 \, \bigr|^{ }_{r = 1}
  \; = \; 
 \frac{
        \kappa^2 v^2 
 \, 
 T \sqrt{m_h^2 - 4 m_\aS^2} 
      }{32\pi^3 \bar{n}^2_\aS}
        \, K^{ }_1\Bigl( \frac{m_h}{T} \Bigr)
 \;. \la{sigmaveq}
\ee

\begin{figure}[t]

\hspace*{-0.1cm}
\centerline{%
 \epsfxsize=7.3cm\epsfbox{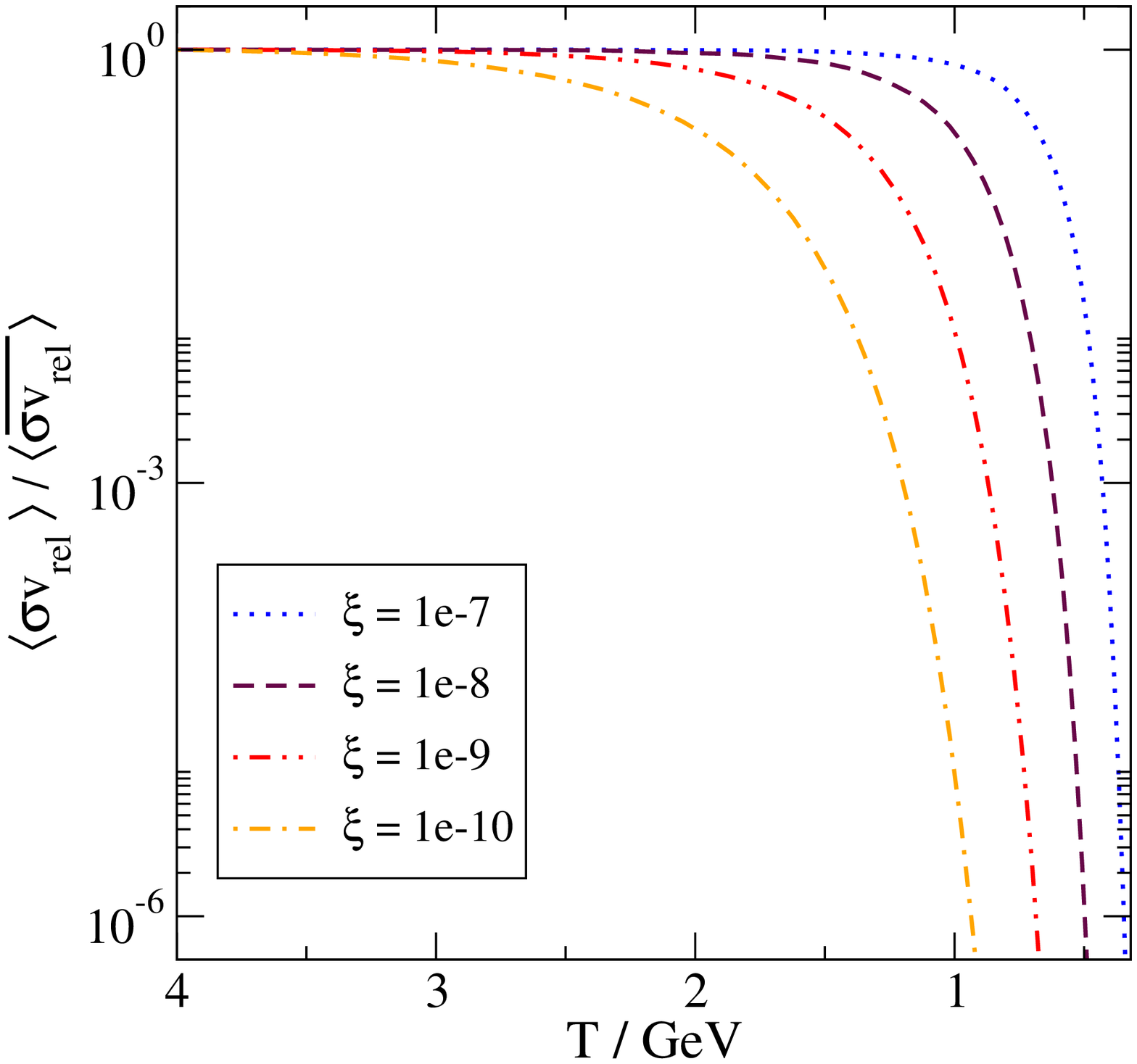}%
 \hspace{0.4cm}%
 \epsfxsize=7.3cm\epsfbox{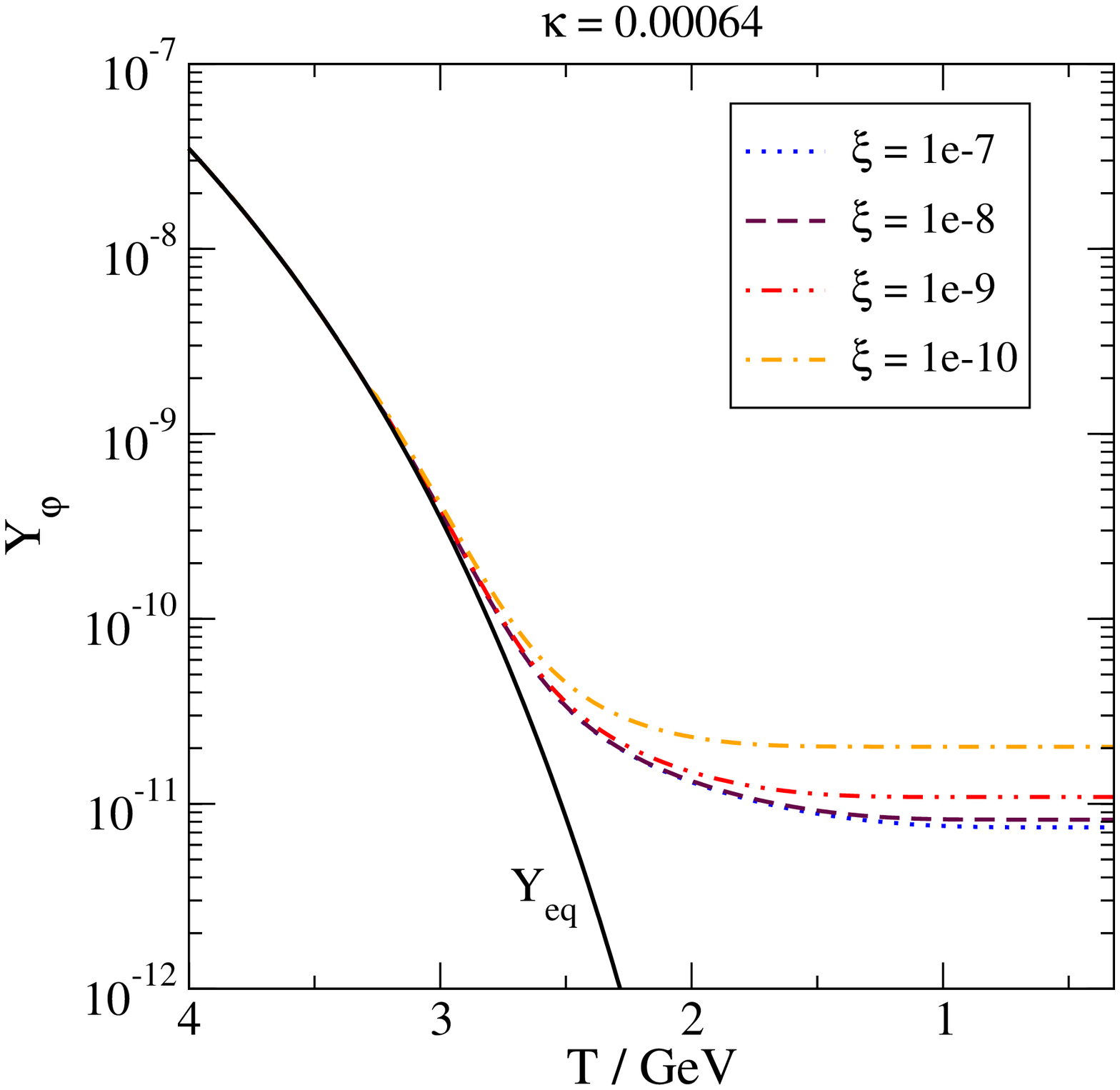}%
}

\caption[a]{\small
  Left:  
  $ \langle \sigma v^{ }_\rmi{rel} \rangle $ from \eq\nr{sigmav}
  compared with the equilibrium  
  $ \langle \overline{ \sigma v }^{ }_\rmi{rel}  \rangle $ 
  from \eq\nr{sigmaveq}, 
  for various values of~$\xi$.
  For small $\xi$, the non-equilibrium 
  momentum distribution is red-tilted
  (cf.\ \fig\ref{fig:tabulate}), having then less weight 
  in the domain that contributes to the 
  annihilation cross section.  
  Right: 
  the corresponding $Y^{ }_\varphi$, 
  obtained from \eq\nr{lw}. Because of the smaller 
  annihilation cross section, the overall yield 
  freezes out earlier, and thus to a larger value. 
  Between $\xi = 10^{-7}_{ }$ and $\xi = 10^{-9}_{ }$, 
  there is a $\sim 45$\% difference in the final yield,
  which could be compensated for by a $\sim 20$\% change of $\kappa$.
  }

\la{fig:sigmav}
\end{figure}

For a numerical illustration, 
we have fixed $m^{ }_{\aS} \simeq 60$~GeV and 
$\kappa \simeq 0.00064$, 
which would yield the correct dark matter abundance 
in kinetic equilibrium according to ref.~\cite{singlet_chemical}. 
However we vary~$\xi$ (cf.\ \eq\nr{xi}), 
in order to obtain an ensemble
of non-equilibrium momentum distributions 
(cf.\ \fig\ref{fig:tabulate}).
The corresponding 
$ 
  \langle \sigma v^{ }_\rmi{rel} \rangle / 
  \langle \overline{ \sigma v }^{ }_\rmi{rel}  \rangle 
$
is shown in \fig\ref{fig:sigmav}(left). 
Inserting into \eq\nr{lw}, 
we obtain the dark matter yield, $Y^{ }_\varphi$,
as illustrated in \fig\ref{fig:sigmav}(right).
As the system falls out of kinetic equilibrium, 
the annihilation cross section is reduced, and 
consequently freeze-out takes place earlier, 
leading to a larger dark matter abundance. 
To keep the dark matter abundance at the correct value, 
the coupling would need to be correspondingly increased, 
however for our benchmark the effect is 
only on the 20~percent level. 

%
\section{Conclusions}
\la{se:concl}

The goal of this paper has been to demonstrate that Langevin simulations
are well suited to studying the efficiency of kinetic equilibration of 
non-relativistic dark matter candidates, produced through 
the freeze-out mechanism. The advantage of the Langevin framework, 
compared with the more standard Boltzmann one, 
is that it cleanly factorizes the slow non-equilibrium problem
from the effect of fast reactions. 
The role of the latter 
is to determine the values of the
matching coefficients in the Langevin description.
The computation of the matching coefficients can be 
viewed as a quantum field theoretic problem, and therefore
pursued up to higher orders of perturbation theory, 
or in principle even non-perturbatively, 
as could be relevant for strong interactions at temperatures
of a few GeV (cf.\ the discussion around \eq\nr{gamma_step2X}).
However, we have remained at the leading order in the current study, 
in order to conform with the accuracy of literature 
studies making use of Boltzmann equations. 

As far as the non-equilibrium problem goes, 
our numerical simulations confirm that the 
momentum distribution retains the Gaussian form despite the 
rapidly evolving matching coefficients 
(that said, the resolution of our setup
is not sufficient for studying momenta in the far UV tail of
the distribution; for this, more advanced techniques
would be required).
Therefore, for practical purposes, it is 
enough to know the width of the momentum distribution, 
given by the quadratic expectation value 
$
 \langle\, \vec{p}^2_{ }\,\rangle
$
(cf.\ \eq\nr{soln_2}).

In order to illustrate these general points, we chose the
example of the scalar singlet model, in a mass regime where 
an efficient $s$-channel resonance drives the freeze-out 
dynamics down to low temperatures. This 
is among the main examples for which the viability of the kinetic
equilibrium assumption has been questioned. 
If the same processes are included in the computation
of the momentum diffusion coefficient (cf.\ \fig\ref{fig:xi}(left))
as in the respective literature~\cite{tb,kk,drake,abe,kk2}, 
our final phenomenological conclusion turns out to be similar. 
In particular, if {\em all} processes are included, then 
$\xi > 10^{-9}_{ }$ in the domain $T > 1$~GeV in which 
the freeze-out dynamics take place (cf.\ \fig\ref{fig:xi}(left)). 
Then kinetic non-equilibrium has a $< 45\%$ influence, 
as shown by a comparison of 
the $\xi = 10^{-9}_{ }$
and $\xi = 10^{-7}_{ }$ curves
in \fig\ref{fig:sigmav}(right), the latter
of which represents practically
the equilibrium solution. 
In terms of the coupling~$\kappa$ (cf.\ \eq\nr{L}),
this corresponds to a $< 20\%$ effect. 

%
\section*{Acknowledgements}

The work of SK is supported by
the National Research Foundation of Korea, 
under grants NRF-2021R1A2C1092701 and NRF-2008-000458,
funded by the Korean government (MEST). The work of ML 
was partly supported by the Swiss National Science Foundation
(SNSF), under grant 200020B-188712.

%
\appendix
\renewcommand{\thesection}{\Alph{section}} 
\renewcommand{\thesubsection}{\Alph{section}.\arabic{subsection}}
\renewcommand{\theequation}{\Alph{section}.\arabic{equation}}

%
\section{Momentum diffusion coefficient in the scalar singlet model}

We illustrate the computation of the momentum diffusion coefficient $\momdif$ 
of \eq\nr{zeta_def}
with the help of the scalar singlet model, defined by the Lagrangian
\be
 \mathcal{L} \; \equiv \; \mathcal{L}^{ }_\rmii{\it SM}
  + \, \biggl\{
     \frac{1}{2} \partial^{\mu}_{ }\varphi \, \partial^{ }_\mu\varphi
  - \, \biggl[
                 \frac{1}{2} \, 
                 \bigl( 
                            m_{\aS 0}^2 + 
                            \kappa\,  \phi^\dagger \phi
                  \bigr) \, \varphi^2
               + \frac{1}{4} \, \lambda^{ }_\aS\, \varphi^4
    \,\biggr]\,  \biggr\} 
 \;, \la{L}
\ee
where $\phi$ is the Higgs doublet. The computation can be 
carried out in a local Minkowskian frame. 
In the Higgs phase, where
$\phi \simeq (0\; v)^T_{ }/ \sqrt{2}$, with $v\simeq 246$~GeV, 
the scalar singlet mass is denoted by $m^{ }_{\aS}$, 
and has the value 
$m^{2}_{\aS} \simeq m_{\aS 0}^2 + \kappa v^2 / 2 $. 
We consider freeze-out dynamics, 
taking place in the regime
$\pi T \ll m^{ }_{\aS} \ll v$. 
Then $\varphi$ can be represented by 
an effective non-relativistic field $\psi$, 
defined as 
\be
 \varphi \simeq \frac{1}{\sqrt{2 m_{\aS}}}
 \,
 \Bigl(\,
   \psi\, e^{- i m^{ }_\aS t } 
   + 
   \psi^*_{ }\, e^{i m^{ }_\aS t } 
 \,\Bigr)
 \;. \la{psi_def}
\ee
Inserting this in \eq\nr{L}, 
integrating over fast oscillations, 
and anticipating the presence 
of a conserved charge (cf.\ \eq\nr{noether}), whereby we 
introduce a chemical potential $\mu \gg T$ guaranteeing
a small overall number density,  
we find that the dynamics of $\psi$
is described by 
\be
 \int \! {\rm d}t \,
 \mathcal{L}^{ } \;\supset\;  
 \int \! {\rm d}t \,
 \psi^*_{ }\,\Bigl\{ 
   i \partial^{ }_0 
 - \mu 
 + \frac{\nabla^2}{2 m^{ }_\aS}
 - \frac{\kappa v h}{2 m^{ }_\aS}
 + \ldots  
 \Bigr\}\, \psi
 \;, \la{non-rel}
\ee
where $h$ is an off-shell mode of the Higgs field, 
with energy $e^{ }_h \ll m^{ }_\aS$. 
Relative to the structures shown, 
the terms omitted
are suppressed by powers of $\partial^{ }_0/m^{ }_{\aS}$, 
$h/v$, 
or $\lambda^{ }_\aS$.

Now, the low-energy description
of \eq\nr{non-rel} has an emergent U(1) symmetry, 
corresponding to conserved particle number.
The Noether current reads
\be
 \mathcal{J}^0_{ } =
 \psi^*_{ }\psi
 \;, \quad
 \mathcal{J}^i_{ } = 
 \frac{\im (\psi^*_{ } \partial^{ }_i \psi)}{m^{ }_\aS} 
 + \ldots 
 \;, \la{noether}
\ee
where the higher-order terms are suppressed by $\nabla^2 / m_\aS^2$.
What is important for us is the ``force'' acting on the 
dark matter particles. By making use of equations of motion
and integrating over the force density, this can be expressed as
\be
 \int_\vec{x} \, 
 \underbrace{ 
   m^{ }_\aS \, \partial^{ }_0 \mathcal{J}^i_{ }
            }_{
                \equiv\; \mathcal{F}^i_\vec{x} 
              }
 = 
 - \frac{\kappa v}{2 m^{ }_\aS}
 \int_\vec{x} 
 \psi^*_{ }\bigl( \partial^{ }_i h \bigr) \psi 
 + 
 \ldots
 \;, \la{F_x}
\ee
where the terms omitted are of the same type 
as in \eqs\nr{non-rel} and \nr{noether}.

In accordance with \eq\nr{zeta_def}, 
the momentum diffusion coefficient $\momdif$ 
is given by the autocorrelator of the force. 
In quantum field theory, 
it is advantageous to define the autocorrelator as the 
zero-frequency limit of a time-symmetrized expectation value, 
\be
 \Gamma[\mathcal{O}^{ }_1,\mathcal{O}^{ }_2]
 \; \equiv \; 
 \lim_{\omega\to 0^+_{ }}
   \int_{-\infty}^{\infty} \! {\rm d}t \, e^{i\omega t}_{ }
   \Bigl\langle \, 
   \frac{1}{2}  
   \bigl\{ \, 
     \mathcal{O}_1^{ }(t)
     \, , \,
     \mathcal{O}_2^{ }(0) 
   \,\bigr\} \, \Bigr\rangle  
   \;. \la{Gamma_def}
\ee
In addition, correlation functions should be normalized so that
the overall density drops out, which can 
be done with the help of the conserved Noether charge, 
\be
 \chi \; \equiv \; 
   \int_\vec{x} 
   \bigl\langle\,
     \mathcal{J}^0_{ }(0^+_{ },\vec{x}) \, 
     \mathcal{J}^0_{ }(0,\vec{0})
   \,\bigr\rangle
 \;, \la{chi_def}
\ee 
where the time can be taken to be Euclidean. 
Thereby the momentum diffusion coefficient 
can be obtained as~\cite{eucl} 
\be
 \momdif \; = \; 
 \frac{
   \frac{1}{3} \sum_{i=1}^3 
   \int_\vec{x} 
   \Gamma[\, 
     \mathcal{F}^i_\vec{x}
     \, , \,
     \mathcal{F}^i_\vec{0} 
         \,]
      }{
     \chi
      }
 \;, \la{gamma_eval}
\ee
where the subscript in $\mathcal{F}^i_\vec{x}$ denotes the spatial position, 
and the definition is from \eq\nr{F_x}. 

In order to evaluate \eq\nr{gamma_eval}, 
the first step is to insert \eq\nr{F_x}, 
go over to a path integral representation, 
and carry out 
the contractions over the 
fields $\psi$ and $\psi^*_{ }$. 
The propagators are non-relativistic (i.e.\ with poles only 
in one half-plane). 
The contractions represent the thermally averaged amplitude squared
for a process in which a heavy $\psi$ interacts with an
off-shell~$h$. Let $\vec{k}$ be the momentum transfer from $h$, 
so that $\vec{q} = \vec{p} + \vec{k}$, where $\vec{p}$ and $\vec{q}$
are the momenta of $\psi$ before and after the interaction. 
Estimating $p\sim\sqrt{m^{ }_\aS T}$ and $k\sim T$
(see below), and employing non-relativistic energies
$\epsilon^{ }_p = p^2 / (2 m^{ }_\aS)$ and 
$\epsilon^{ }_q = q^2 / (2 m^{ }_\aS)$, 
the Boltzmann weight does not depend on $k$ 
to leading order in $T/m_\aS^{ }$, 
\be
 \frac{\epsilon^{ }_q - \epsilon^{ }_p}{T}
 \;\sim\; 
 \frac{\vec{p}\cdot\vec{k}}{m^{ }_\aS T}
 \;\lsim\; 
 \sqrt{\frac{T}{m^{ }_\aS}}
 \;\ll\; 
 1
 \;. 
\ee
Therefore the thermal average over $\vec{k}$ effectively localizes 
$h^{ }_\vec{x}$, 
\be
 \int_\vec{k} 
 \int_\vec{x} e^{i\vec{k}\cdot\vec{x}}_{ } \,
 h^{ }_\vec{x} = h^{ }_\vec{0}
 \;. 
\ee
Furthermore the overall density of the charge matter 
particles (determined by the chemical potential $\mu$)
cancels between the numerator and denominator. 
Thus we are left with
\be
 \momdif \; = \; 
 \frac{\kappa^2 v^2}{12 m_\aS^2 }
 \sum_{i=1}^{3}
 \Gamma[\, 
   \partial^{ }_i h^{ }_\vec{0}
   \,,\,
   \partial^{ }_i h^{ }_\vec{0}
 \,]
 \; + \; \ldots
 \;. \la{gamma_step1}
\ee

The next step is to consider various interactions
experienced by the Higgs boson, 
\be
 S^{ }_\rmii{I} \; \supset \; 
 \int_{\X = (t,\vec{x})} h(\X)\, \mathcal{O}(\X) 
 + \ldots
 \;, 
\ee
where only operators linear in $h$ need to be included at leading
order. 
Going to momentum space and 
contracting over the (off-shell) Higgs boson yields
\be
 \momdif \; = \; 
 \frac{\kappa^2 v^2}{12 m_\aS^2 }
 \int_\vec{k} 
 \frac{k^2
      }{(k^2 + m_h^2)^2}
 \int_\vec{x} e^{i\vec{k}\cdot\vec{x}}_{ }\,
   \Gamma[\,
   \mathcal{O}^{ }_{\vec{x}}
   \,,\,
   \mathcal{O}^{ }_{\vec{0}}
   \,]
 \; + \; \ldots
 \;, \la{gamma_step2}
\ee
where 
$
 k \equiv |\vec{k}|
$,
$
 \int_\vec{k} = 
 \int \! \frac{{\rm d}^3\vec{k}}{(2\pi)^3_{ }}
$, 
and 
we went to the static limit as required by
the definition of $\Gamma$.

A further simplification follows by noting that
the momentum integral is saturated by $k\sim \pi T \ll m^{ }_h$,  
cf.\ \eqs\nr{eta_F} and \nr{eta_B}. Therefore 
the momentum diffusion coefficient can be approximated as 
\be
 \momdif \; \approx \; 
 \frac{\kappa^2 v^2}{12 m_\aS^2 m_h^4}
 \int_\vec{k} 
 k^2  \!\!
 \int_\vec{x}
 \,  
 e^{i\vec{k}\cdot\vec{x}}_{ }
 \, 
   \Gamma[\,
   \mathcal{O}^{ }_{\vec{x}}
   \,,\,
   \mathcal{O}^{ }_{\vec{0}}
   \,]
  \; + \; \ldots
 \;. \la{gamma_step2X}
\ee
In other words, the Higgs exchange is a contact interaction
at low energies. 
We stress that the operator $  \mathcal{O}^{ }_{\vec{x}} $
is gauge invariant under QCD, and that therefore 
\eq\nr{gamma_step2X} is defined and 
computable beyond perturbation theory. 

To proceed with a leading-order evaluation 
of \eq\nr{gamma_step2X}, consider first 
a fermionic operator containing the $b$ quark, 
\be
 \mathcal{O}^{ }_\F \; \equiv \; 
 - \frac{h^{ }_b\, \bar{b} b }{\sqrt{2}} 
 \;. \la{O_F_def}
\ee
In the numerical illustrations in \fig\ref{fig:xi}, 
we also include the $c$ quark. 
A few-page thermal field theory computation yields  
\be
 \int_\vec{x}
 \,  
 e^{i\vec{k}\cdot\vec{x}}_{ }
 \, 
 \Gamma[\,
   \mathcal{O}^{ }_{\F,\vec{x}}
   \,,\,
   \mathcal{O}^{ }_{\F,\vec{0}}
 \,]
 \; = \;
 \frac{h_b^2 \Nc^{ }(k^2 + 4 m_b^2) T}{4\pi k}
 \, \nF^{ }
 \Bigl(\, 
    \sqrt{k^2/4 + m_b^2} 
 \,\Bigr) 
 + \rmO(h_b^2 \alphas^{ })
 \;, \la{eta_F}
\ee
where $\nF^{ }$ is the Fermi distribution. 
Changing variables, we finally get
\be
 \momdif^{ }_\F 
 \; \approx \; 
 \frac{4 \kappa^2 m_b^2 \Nc^{ } T}{3\pi^3 m_\aS^2 m_h^4 }
 \int_{m^{ }_b}^\infty \! {\rm d}y\, y^3 (y^2 - m_b^2 ) \,\nF^{ }(y)
 \; \stackrel{ 
             }{\le} \;
 \frac{31 \pi^3 \kappa^2 m_b^2 \Nc^{ }T^7_{ }}
      {189 m_\aS^2 m_h^4}
 \;, \la{gammaF_est} 
\ee
where the upper bound is saturated in the limit $m^{ }_b \ll \pi T$.
Numerically, the upper bound 
gives a fairly good approximation for charm quarks.

As a second example, we consider the bosonic operator obtained
by integrating out the top quark~\cite{dim5},
\be 
 \mathcal{O}^{ }_\B \; \equiv \; 
 - \frac{ \alphas^{ } G^a_{\mu\nu} G^{a\mu\nu}_{ } }{12\pi v}
 \;, 
\ee 
whose relevance for thermal considerations has been 
underlined in ref.~\cite{gw}. 
In this case we obtain
\be
 \int_\vec{x}
 \,  
 e^{i\vec{k}\cdot\vec{x}}_{ }
 \, 
 \Gamma[\,
   \mathcal{O}^{ }_{\B,\vec{x}}
   \,,\,
   \mathcal{O}^{ }_{\B,\vec{0}}
 \,]
 \; = \;
 \biggl( \frac{\alphas^{ }}{12\pi v} \biggr)^2_{ }
 \frac{(\Nc^2 -1) k^3_{ } T}{\pi}
 \, \nB^{ }
 \Bigl(\, \frac{k}{2} \,\Bigr) 
 + \rmO(\alphas^3)
 \;, \la{eta_B}
\ee
where $\nB^{ }$ is the Bose distribution. Changing variables and
carrying out the integral yields 
\be
 \momdif^{ }_\B  
 \; \approx \;
 \frac{2 \kappa^2 \alphas^2 (\Nc^{2} -1) T}{27 \pi^5 m_\aS^2 m_h^4 }
 \int_{0}^\infty \! {\rm d}y\, y^7_{ } \nB^{ }(y)
 \; = \; 
 \frac{16 \pi^3 \kappa^2 \alphas^2 (\Nc^{2} - 1) T^9_{ }}
      {405 m_\aS^2 m_h^4}
 \;. \la{gammaB_est} 
\ee

In order to illustrate the magnitude of these corrections, 
we show in \fig\ref{fig:xi}(left) how they can be converted into
the coupling $\xi$, defined in \eq\nr{xi}. 

\small{
%

}

\end{document}